\documentclass[floatfix,aps,twocolumn,showpacs,superscriptaddress,amsmath,amssymb,longbibliography]{revtex4-2}  
 
\usepackage[autostyle=true]{csquotes}
\usepackage{times}
\usepackage{amsmath}
\usepackage{mathtools}
\usepackage{textcomp}
\usepackage{gensymb}
\usepackage{graphicx}
\usepackage{siunitx}
\usepackage{ulem}
\usepackage{physics}
\usepackage{appendix}
\newcommand{\dbar}{{\mkern3mu\mathchar'26\mkern-12mu d}}
\usepackage{ulem}
\usepackage[unicode=true,
 bookmarks=true,bookmarksnumbered=true,bookmarksopen=true,bookmarksopenlevel=2,
 breaklinks=false,pdfborder={0 0 1},backref=false,colorlinks=true]
 {hyperref}
 \hypersetup{linkcolor=blue, citecolor=blue, urlcolor=blue, filecolor=blue, pdfpagelayout=OneColumn,
  pdfnewwindow=true, pdfstartview=XYZ, plainpages=false}

\usepackage[all]{hypcap}
\usepackage{color}

\makeatletter
\def\mathcolor#1#{\@mathcolor{#1}}%
\def\@mathcolor#1#2#3{%
  \protect\leavevmode%
  \begingroup\color#1{#2}#3\endgroup%
}%
\newcommand{\msout}[1]{\text{\sout{\ensuremath{#1}}}}%

\newcommand{\sam}[2]{%
\ifmmode%
  \msout{#1}\mathcolor{red}{#2}%
\else%
  \sout{#1}\textcolor{red}{#2}%
\fi}
\newcommand{\samO}[2]{%
\ifmmode%
  \msout{#1}\mathcolor{magenta}{#2}%
\else%
  \sout{#1}\textcolor{magenta}{#2}%
\fi}

\newcommand{\cyr}[2]{%
\ifmmode%
  \msout{#1}\mathcolor{red}{#2}%
\else%
  \sout{#1}\textcolor{red}{#2}%
\fi}

\newcommand{\gab}[2]{%
\ifmmode%
  \msout{#1}\mathcolor{green}{#2}%
\else%
  \sout{#1}\textcolor{green}{#2}%
\fi}

\makeatother

\makeatletter

\newcommand\Autoref[1]{\@first@ref#1,@}
\def\@throw@dot#1.#2@{#1}
\def\@set@refname#1{
    \edef\@tmp{\getrefbykeydefault{#1}{anchor}{}}%
    \xdef\@tmp{\expandafter\@throw@dot\@tmp.@}%
    \ltx@IfUndefined{\@tmp autorefnameplural}%
         {\def\@refname{\@nameuse{\@tmp autorefname}s}}%
         {\def\@refname{\@nameuse{\@tmp autorefnameplural}}}%
}
\def\@first@ref#1,#2{%
  \ifx#2@\autoref{#1}\let\@nextref\@gobble
  \else%
    \@set@refname{#1}
    \@refname~\ref{#1}
    \let\@nextref\@next@ref
  \fi%
  \@nextref#2%
}
\def\@next@ref#1,#2{%
   \ifx#2@ and~\ref{#1}\let\@nextref\@gobble
   \else, \ref{#1}
   \fi%
   \@nextref#2%
}

\makeatother

\begin{document}

\def\be{\begin{equation}}
\def\ee{\end{equation}}

\def\myvec#1{{\bf #1}}
\def\Esw{\myvec{E}_{sw}}
\def\Hsw{\myvec{H}_{sw}}

\title{Optimal time-entropy bounds and speed limits for Brownian thermal shortcuts}

\author{Lu\'{\i}s Barbosa Pires}
\affiliation{University of Strasbourg and CNRS, CESQ and ISIS, UMR 7006, F-67000 Strasbourg, France}

\author{R\'emi Goerlich}
\affiliation{University of Strasbourg and CNRS, Institut de Physique et Chimie des Mat\'eriaux de Strasbourg, UMR 7504, F-67000 Strasbourg, France}
\affiliation{University of Strasbourg and CNRS, CESQ and ISIS, UMR 7006, F-67000 Strasbourg, France}

\author{Arthur Luna da Fonseca}
\affiliation{Instituto de Fisica, Universidade Federal do Rio de Janeiro, Caixa Postal 68528, Rio de Janeiro, RJ, 21941-972, Brazil}
\affiliation{University of Strasbourg and CNRS, CESQ and ISIS, UMR 7006, F-67000 Strasbourg, France}

\author{Maxime Debiossac}
\affiliation{Vienna Center for Quantum Science and Technology, Faculty of Physics, University of Vienna, A-1090 Vienna, Austria}

\author{Paul-Antoine Hervieux}
\affiliation{University of Strasbourg and CNRS, Institut de Physique et Chimie des Mat\'eriaux de Strasbourg, UMR 7504, F-67000 Strasbourg, France}

\author{Giovanni Manfredi}
\email{giovanni.manfredi@ipcms.unistra.fr}
\affiliation{University of Strasbourg and CNRS, Institut de Physique et Chimie des Mat\'eriaux de Strasbourg, UMR 7504, F-67000 Strasbourg, France}

\author{Cyriaque Genet}
\email{genet@unistra.fr}
\affiliation{University of Strasbourg and CNRS, CESQ and ISIS, UMR 7006, F-67000 Strasbourg, France}

\date{\today}

\begin{abstract} 

By controlling in real-time the variance of the radiation pressure exerted on an optically trapped microsphere, we engineer temperature protocols that shortcut thermal relaxation when transferring the microsphere from one thermal equilibrium state to an other. We identify the entropic footprint of such accelerated transfers and derive optimal temperature protocols that either minimize the production of entropy for a given transfer duration or accelerate as much as possible the transfer for a given entropic cost. Optimizing the trade-off yields time-entropy bounds that put speed limits on thermalization schemes. We further show how optimization expands the possibilities for accelerating Brownian thermalization down to its fundamental limits. Our approach paves the way for the design of optimized, finite-time thermodynamic cycles at the mesoscale. It also offers a platform for investigating fundamental connections between information geometry and finite-time processes.

\end{abstract}

\maketitle

The time needed for a body to thermalize with its environment is a natural constraint for operating many physical systems and devices. Controlling thermalization has emerged as one salient challenge at meso and nanoscales scales \cite{chang2009cavity,martinez2017colloidal,albay2021shift,gonzalez2021levitodynamics,rademacher2022nonequilibrium}. At such scales, the methods of stochastic thermodynamics have proven their efficiency, capable of extending the concepts of work, heat and entropy to single, fluctuating systems \cite{sekimoto2010stochastic,seifert2012stochastic}. Experimentally, new strategies have recently been implemented on optically trapped Brownian particles to emulate effective, and thereby controllable, thermal baths \cite{raizen2012optical,martinez2013effective,chupeau2018thermal,delic2020cooling,van2021sub}. The fine control of the time-dependence of effective temperatures has led to the definition of thermal protocols and optimized cycles \cite{martinez2016brownian,plata2020building,watanabe2022finite}. Exploiting finite-time thermodynamics, these strategies have also provided the means to circumvent natural thermalization by proposing accelerated paths that a Brownian system can be forced to follow \cite{chupeau2018thermal,kumar2020exponentially,plata2020finite,nakamura2020fast,jun2021instantaneous,
chen2022optimizing,patron2022thermal}. Such means form a major topic of current research in the realm of shortcuts to adiabaticity \cite{guery2019shortcuts, guery2022driving}.

Obviously, speeding-up transitions from one equilibrium state to another demands to follow non-equilibrium paths that have a thermodynamic cost. Once such cost evaluated, the design of protocols that optimize the mutually exclusive relation between the rate of acceleration and the energetic expense should be possible. 
There is a variety of approaches proposed for evaluating that energetic expense \cite{albay2019thermodynamic,debiossac2020thermodynamics,prados2021optimizing,frim2022optimal,ye2022optimal,jun2022minimal,paraguassu2022effects}, but the challenge remains to identify the proper one which makes it possible to treat duration and cost on an equal footing, the prerequisite for this optimization \cite{schmiedl2007optimal,deffner2018minimal,rosales2020optimal}.

In this Letter, we set up a bath engineering strategy involving radiation pressure to directly control the kinetic temperature of an optically trapped, overdamped, Brownian microsphere \cite{goerlich2022harvesting}. This control allows us to impose abrupt transfers from one to another equilibrium states, either increasing or decreasing the temperature down to a minimum set by room temperature $T_R$. Such step-like transfers are followed by thermal relaxations measured precisely through the diffusive dynamics of the microsphere inside the harmonic optical trap. Our strategy gives the possibility to accelerate such thermal relaxation processes by imposing an overshoot in temperature during the transfer. This leads us to extend to isochoric transitions the Engineered Swift Equilibration (ESE) processes developed so far for isothermal transitions \cite{martinez2016engineered}. We show how thermal ESE protocols -- hereafter named ThESE protocols -- do accelerate thermalization, and demonstrate experimentally the shortening of the duration of initial-to-final thermal equilibrium transfers. We further quantify the thermodynamic cost of this acceleration in terms of entropy production.

The identification of the entropic cost of a thermal shortcut brings us, in the context of harmonic trapping, to a class of optimal thermal protocols, defined as those protocols that speed-up thermalization while minimizing the associated production of entropy. The trade-off involved in our optimization procedure sets time-entropy bounds that put speed limits on physically realizable thermal shortcuts, in a remarkable asymmetry between heating and cooling protocols. We also demonstrate that optimal cooling gives access to higher acceleration rates that are unreachable using standard overshooting, ThESE-like protocols. These results are finally discussed from an energetic viewpoint for the three families (step-like, ThESE and optimized) of state-to-state transitions, which clarifies the different contributions to the global in-take of heat by the trapped microsphere under the action of the fluctuating radiation pressure. We also stress that our optimization-under-cost constraint leads to results that are different from the thermal brachistochrones recently proposed in \cite{prados2021optimizing}.

Our experiment consists of a single microsphere trapped in an optical tweezer and evolving in a harmonic potential~\cite{padgett2010optical, gennerich2017optical}. The microsphere diffuses in water with a Stokes drag $\gamma=2.695 \times 10^{-8}$ kg/s at room temperature $T_R=293 K$. The trap is characterized by a stiffness $\kappa=13.1 \pm 0.2$ fN/nm and the overdamped diffusion dynamics by a relaxation time $\tau = \gamma/\kappa=2.06\pm 0.04$ ms. As described in detail in Appendices \ref{app.setup} and \ref{app.calibration}, an additional radiation pressure is exerted on the sphere by a pushing laser whose intensity $I(t)= I_{0}+\delta I(t)$ is digitally controlled over time by an acousto-optic modulator \cite{goerlich2022harvesting}. When $\delta I(t)$ is random (white noise spectrum), this radiation pressure increases the motional variance of the center-of-mass motion of the sphere along the optical axis of the trap. By building a statistical ensemble $\{j\}$ of $1.7\times 10^4$ diffusing trajectories $x_j(t)$, we extract an ensemble average variance $s(t)$ in direct relation with the pushing laser intensity variance $\langle\delta I^{2}(t)\rangle$. 
This random forcing of the microsphere can be interpreted as emulating an effective thermal bath whose temperature $T(t)$ can be set instantaneously in strict relation with the intensity variance with $T(t)\propto\langle\delta I^{2}(t)\rangle$. The mechanical response of the microsphere is measured through the time-evolution of $s(t)$ according to --see Appendix \ref{APPENDIX_ExtNoise}:
\begin{equation}
\frac{d s(t)}{dt}= \frac{2}{\tau}\left(\frac{k_{\rm B} T(t)}{\kappa}-s(t) \right).  \label{eq:var2}
\end{equation}

The effective nature of $T(t)$ implies that thermal changes impact the diffusion coefficient simply like $D(t)=k_{\rm B}T(t)/\gamma$ and the system relaxation time $\tau$ remains constant. By thus transforming temperature into an external control parameter $T(t)$, the crucial asset of our bath engineering strategy is the possibility to perform specific temperature \textit{protocols} that can be arbitrarily fast from one initial $T_R+T_i$ to another final target temperature $T_R+T_f$. As discussed below, this opens rich analogies with recent works that have demonstrated how time-dependent optical trap stiffness protocols $\kappa(t)$ can lead to shortcut, engineer and even optimize state-to-state isothermal processes \cite{martinez2016engineered,lecunuder2016fast,rosales2020optimal}. 

\begin{figure}[htb!]
	\centering{
		\includegraphics[width=0.4\textwidth]{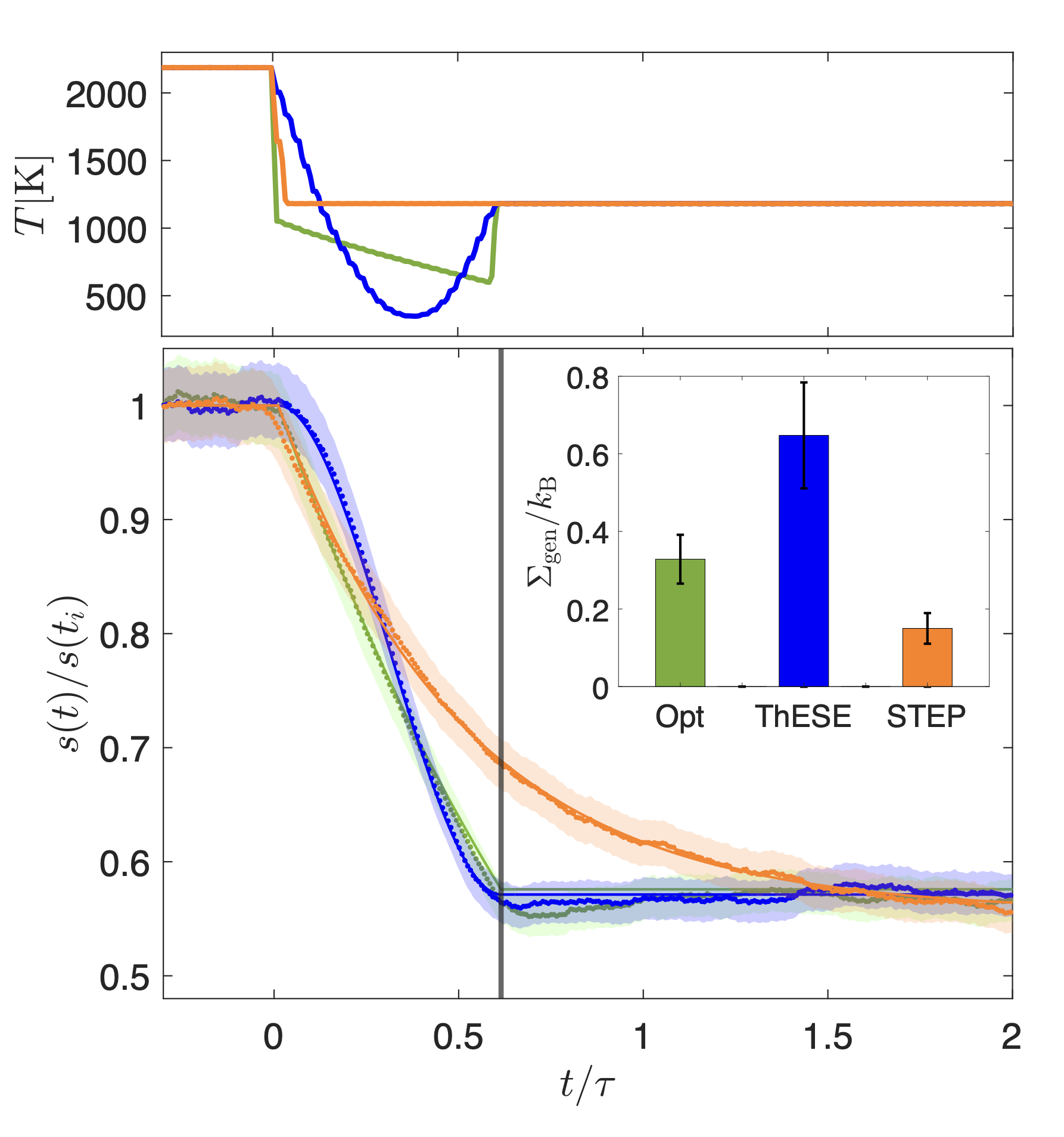}}
	\caption{Upper panel: Time evolution $T(t)$ plotted for three cooling protocols connecting two thermal equilibrium states, one initial at $T_R+T_i$ and one final at $T_R+T_f$ where $T_R=300$ K corresponds to room temperature and $T_R+T_i=2200$ K, $T_R+T_f=1200$ K to the target temperatures fixed by the laser fluctuation spectrum. The orange curve corresponds to a step-like protocol (STEP) with a simple temperature discontinuous quench, the blue curve to the ThESE protocol through which $T_{i}$ and $T_{f}$ are connected by a third degree polynomial, the green curve to the optimal protocol (Opt) with its two discontinuities ensuring the connection between the initially hot and finally cold equilibrium states. Lower panel: Measured time evolutions (data points with same color coding as in the upper panel) of the motional variances $s(t)$ induced by the different thermal protocols starting at $t_i/\tau=0$. The variances for the ThESE and Opt shortcuts reach equilibrium at a rate $\Delta t/\tau = 0.6$, as indicated by the vertical line. The shaded areas give the experimental errors at a $95\%$ confidence interval for $s(t)$. The analytical solutions for $s(t)$ for the three protocols, calculated in Appendix \ref{app:dimen} from Eq. (\ref{eq:var2}), correspond to the continuous lines drawn on the experimental data points. Inset: Levels of produced entropy $\Sigma_\text{gen}$ associated to each of the three protocols and evaluated using Eq. (\ref{eq:Sgen}). Error bars for $\Sigma_\text{gen}$ correspond to the combination of the uncertainty in $s(t)$, stiffness and temperature (see more details in Appendix \ref{app:error}).}
	\label{fig1}
\end{figure}

Let us start by implementing a sudden, step-like change $\Delta T=T_f-T_i$ between an initial $T_R+T_i$ to a final $T_R+T_f$ temperature. This STEP protocol is described in the upper panel of Fig. \ref{fig1}. It produces a transient response of the variance $s(t)$ that we measure and plot in the lower panel of  Fig. \ref{fig1}. As observed in excellent agreement with Eq. (\ref{eq:var2}), the variance relaxes towards the new thermal equilibrium state with a relaxation time $2\tau$. Such a relaxation corresponds to the definition of \textit{natural} thermalization, in which the system is left free to evolve towards a new equilibrium state. We now show that it is possible to impose a temperature protocol that displays a much shorter thermalization time for the same $\Delta T$ change. To do so, we extend to temperature the class of ESE isothermal protocols presented in \cite{martinez2016engineered} with a polynomial temperature overshoot evaluated in Appendix \ref{app:these}, imposing stationary thermal equilibrium $\kappa s(t_{i,f})=k_{\rm B}(T_R+T_{i,f})$ and $\dot{s}(t_{i,f})=0$ at the initial and final steps of the process. The corresponding $T(t)$ protocol is plotted in the upper panel of Fig. \ref{fig1} for a chosen transfer duration $\Delta t=t_f-t_i = 1.23$ ms imposed to be shorter than $2\tau$ with a ratio $\Delta t / \tau = 0.6$. It is implemented experimentally and we measure in the lower panel of Fig. \ref{fig1} the time evolution of the system's variance $s(t)$ in excellent agreement with the theory.

The central question for such accelerated thermalization protocols remains their possible optimization with respect to a well-identified footprint. In the case of isothermal stiffness protocols $\kappa (t)$, the thermodynamic cost of acceleration was evaluated through the associated work expense and the minimization procedure designed accordingly \cite{schmiedl2007optimal,deffner2018minimal,rosales2020optimal}. Temperature protocols, in contrast, are entropic by nature and have been recently characterized using the concept of thermal (entropic) work \cite{rademacher2022nonequilibrium}. This entropic nature clearly appears when interpreting Eq. (\ref{eq:var2})  thermodynamically: whenever the temperature changes faster than $\tau$, the system will evolve along an irreversible, non-equilibrium process in which the instantaneous variance $s(t)$ will be different from the one expected by equipartition. The difference between $s(t)$ and $k_{\rm B}T(t)/\kappa$ thus measures the deviation from a reversible process and, as such, is associated to a given production of entropy that we now evaluate.

For our experiments, we define the system's stochastic entropy $\sigma_{\rm sys}(x_j(t),T(t))=-k_{\rm B} \ln p(x_j(t),T(t))$ \cite{seifert2012stochastic} from an extension of the Boltzmann probability density $p(x_j(t),T(t))=\sqrt{\kappa/2\pi k_{\rm B}T(t)} \exp ( -\kappa x^2_j(t)/(2 k_{\rm B}T(t)))$ to non-equilibrium processes that connect two equilibrium states. Using this definition, the infinitesimal variation of the system's entropy is evaluated as $d\sigma_{\rm sys}(x_j(t),T(t))=\kappa x_idx_j/T(t) +(k_{\rm B}T(t)-\kappa x_j^2)dT/(2T^2(t))$. The first term involves the quantity of heat $dq=-\kappa x_idx_j$ associated with the change of internal of energy of the system. It corresponds to (the opposite of) the variation of the entropy of the medium $\dbar \sigma_{\rm med}=dq/T(t)$ where we use the $\dbar$ notation for non-exact differentials. The second term, written as $d\sigma_{\rm sys}+\dbar \sigma_{\rm med}$, thus gives the infinitesimal amount of total entropy generated along the elementary path $dT$ --see Appendix \ref{app:Generation_of_entropy}.

The generated total entropy, once ensemble averaged and cumulated from the initial $t_i$ to a given time $t$ of the isochoric transformation
\begin{equation}\label{eq:Sgen}
\Sigma_{\rm gen}(t)=\frac{1}{2} \int_{t_i}^{t}\frac{\dot{T}(\zeta)}{T^2(\zeta)}\left(k_{\rm B}T(\zeta)-\kappa s(\zeta)\right)d\zeta,
\end{equation}
directly involves the non-equilibrium nature of the transformation imprinted in the difference $k_{\rm B}T(t)-\kappa s(t)$ between the measured variance and equipartition. It therefore corresponds to the cumulated entropy produced along the irreversible transition and constitutes the entropic footprint of a finite-time isochoric process.
For the entire STEP and ThESE protocols, in which $t\rightarrow\infty$ and $t=t_f$ respectively, the total entropy production can be easily calculated using Eq. (\ref{eq:Sgen}), and the results are plotted in the inset of Fig. \ref{fig1}. When compared, these values reveal in a striking manner the entropic cost of thermal acceleration with $\Sigma_{\rm gen}^{\rm ThESE}>\Sigma_{\rm gen}^{\rm STEP}$. 

Our analysis now gives the possibility to derive the actual temporal profile of an optimal protocol that minimizes this entropic cost for a given choice of transfer duration $\Delta t=t_f-t_i$ from one equilibrium state at $T_R+T_i$ to another at $T_R+T_f$. In close relation with our previous work \cite{rosales2020optimal}, we express the transfer duration as a functional of the variance according to Eq. (\ref{eq:var2}) and integrate by parts the generated entropy (\ref{eq:Sgen}) --see Appendix \ref{app:optimalprot}-- to build a functional
\begin{equation}
\label{eq.euler-lagrange}
J[T(s)]  = \int_{s_{i}}^{s_{f}} \left(  \frac{\gamma}{k_{\rm B}T(s) -s\kappa} - \lambda \frac{\kappa}{k_{\rm B}T(s)} \right)\textrm{d} s.
\end{equation}
This combines on an equal footing the transfer duration and the corresponding generation of entropy with a Lagrange multiplier $\lambda$ to regulate the trade-off between the two quantities. The optimization procedure consists in searching for the paths in the $[s,T(s)]$ space that minimize $J[T(s)]$ while keeping the same initial and final equilibrium conditions imposed by the equipartition theorem, just like for the ThESE protocol. As explained in Appendix \ref{app:dimen}, the procedure yields two families of optimized thermal protocols $T_{\rm heat/cool}(s)$ associated respectively to heating $T_i<T_f$ and cooling $T_i>T_f$. We emphasize that the $T_{\rm heat/cool}(s)$ protocol does not satisfy thermal equilibrium at both initial and final times and must be therefore supplemented by two discontinuous transitions, just like in the case of optimal isothermal processes \cite{schmiedl2007optimal, rosales2020optimal}. During the interval $\Delta t$, those two solutions correspond to an exponential evolution of the variance with $s_{\rm opt}(t) = s_i ((T_R+T_f)/(T_R+T_i))^{(t-t_i)/\Delta t}$ . It is remarkable that both $T_{\rm heat/cool}(s)$ protocols can be described with a single expression
\begin{equation}
\label{eq:sol.cool.expli2}
T_{\rm opt}(t) = \frac{\kappa s(t)}{k_{\rm B}}\left( 1+\frac{\tau}{ k_{\rm B}}\frac{\Delta \Sigma_{\rm sys}}{\Delta t} \right)
\end{equation}
with $\Delta\Sigma_{\rm sys}=k_{\rm B}\ln [ (T_R+T_f)/(T_R+T_i)]/2$ the protocol-independent total variation in the system's entropy.
This expression is plotted in the upper panel of Fig. \ref{fig1} for an optimal cooling protocol and with the same shortening rate $\Delta t / \tau = 0.6$ as the ThESE protocol discussed above. As one expected important result of our work, $\Sigma_{\rm gen}^{\rm ThESE}>\Sigma_{\rm gen}^{\rm opt}$. 

The optimal protocol given by Eq. (\ref{eq:sol.cool.expli2}) together with $s_{\rm opt}(t)$ injected into Eq. (\ref{eq:Sgen}) lead to evaluate the minimal entropy produced through an isochoric transformation of duration $\Delta t$ as
\begin{equation}
\label{eq:Smin}
 \Sigma_{\rm min} =  \frac{\Delta \Sigma_{\rm sys}}{1+\frac{k_{\rm B}}{\tau}\frac{\Delta t}{\Delta \Sigma_{\rm sys}}},
\end{equation}
an expression valid both for cooling and heating protocols but with different consequences, as discussed below.

\begin{figure}[htb!]
	\centering{
		\includegraphics[width=0.4\textwidth]{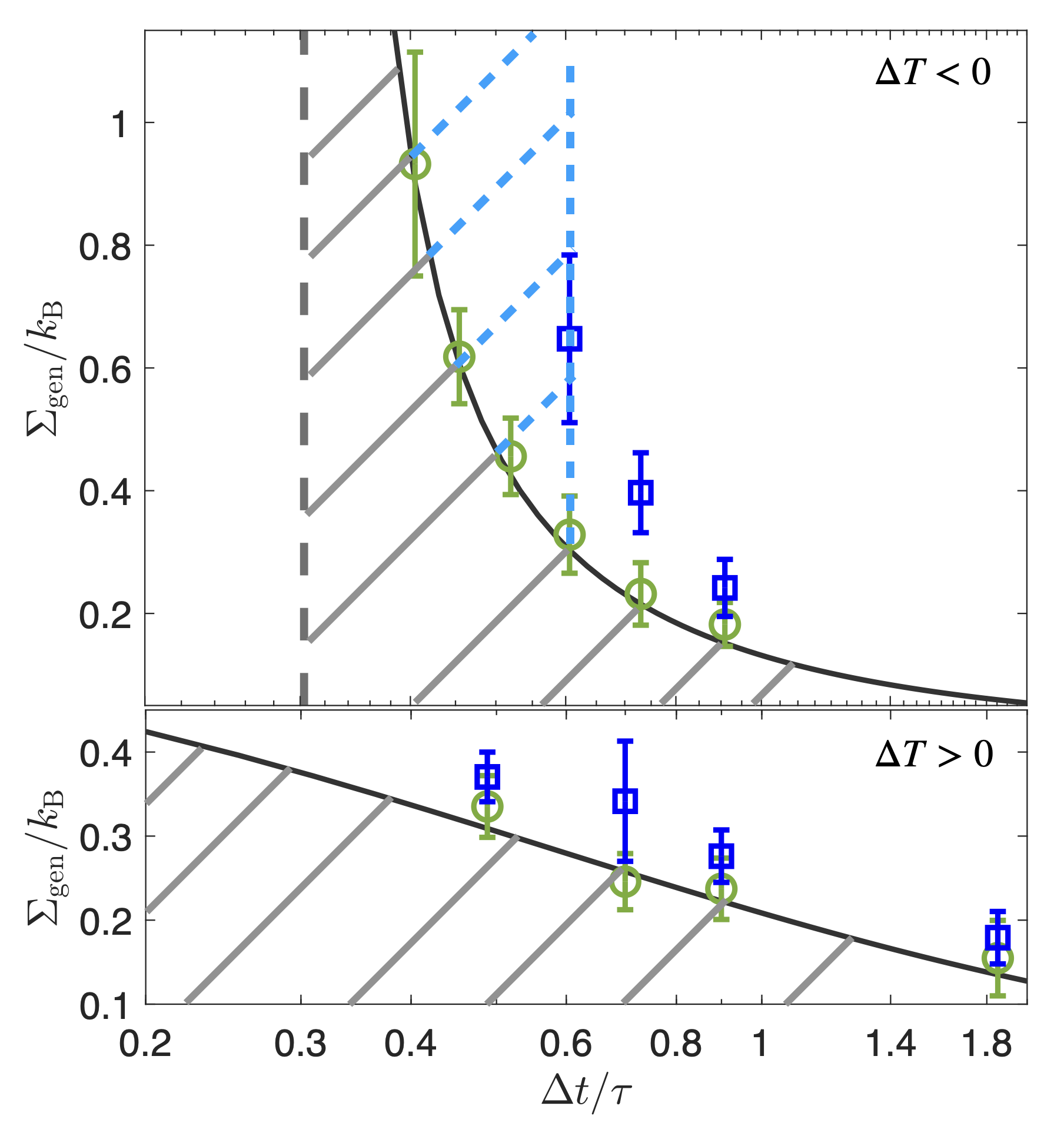}}
	\caption{Minimal time-entropy bound (solid black lines) corresponding to an optimal $(T_R+T_i= 2200 {\rm K}, T_R+T_f=1200 {\rm K})$  cooling process (upper panel) and heating protocols performed between $T_R+T_i=350$ K to $T_R+T_f=1100$ K (lower panel). The gray hatched region are forbidden to any acceleration method. Experimental measurements for optimal protocols are shown with green open circles and with blue open squares for ThESE protocols. Error bars correspond to experimental errors propagated through Eq. (\ref{eq:Sgen}) at a $95\%$ confidence level. The fundamental limit put on ThESE cooling protocols set at $\Delta t/\tau = 0.6$ for the chosen experimental parameters is depicted as a second exclusion region (blue hatched region) for such overshooting temperature protocols. }
	\label{fig2}
\end{figure}

In Fig. \ref{fig2}, we first plot Eq. (\ref{eq:Smin}) for cooling (upper panel) and heating (lower panel) optimal protocols (solid black lines). The curves draw exclusion regions for entropy production that correspond to the minimal amount of entropy that can be generated in an isochore for a given $\Delta t$: they thus correspond to optimal time-entropy bounds. Our experimental results obtained for different optimal cooling and heating protocols $T_{\rm opt} (t)$ injected within our optical trap (same set of temperatures but different transfer durations) all precisely fall on the expected bounds. 

For a cooling process with $\Delta \Sigma_{\rm sys}<0$, Eq. (\ref{eq:Smin}) also puts an asymptotic limit to the transfer rate with a minimal transfer duration of $\Delta t_{\rm min}/\tau = -\Delta \Sigma_{\rm sys}/k_{\rm B} $. These bounds on the dynamical evolution of our system, extracted from the trade-off involved in the optimization procedure between the transfer duration and the production of entropy, must be considered as true speed limits on the state-to-state connection \cite{shiraishiSpeed2018}. They are directly associated with a divergence in the entropic cost as clearly seen experimentally in Fig. \ref{fig2} for the shortest transfer rate that we probed (vertical dashed line). This limit in the cooling acceleration is directly related to the fact that the lowest temperature $T_{\rm min}=T_{\rm opt} (t_f^-)=T_f (1+\frac{\tau}{k_{\rm B}}\frac{\Delta \Sigma_{\rm sys}}{\Delta t})$ which the optimal protocol passes through, cannot be smaller than $0$ K, a temperature limit reached when $\Delta t_{\rm min}/\tau = -\frac{\Delta \Sigma_{\rm sys}}{k_{\rm B}} =0.3$ for our $(T_i,T_f)$ choice. However, experimentally we necessarily have $T_{\rm min}\geq T_R$ and for the case presented in Fig. \ref{fig2}, this implies that the shortest transfer rate reachable is $\Delta t_{\rm min}/\tau=\frac{\Delta \Sigma_{\rm sys}}{k_{\rm B}(1-T_R/T_f)}\simeq 0.4$. 

Room temperature obviously bounds from below all overshoot temperatures that can be physically hit. This leads to an interesting consequence when comparing optimal and ThESE cooling protocols for identical shortening rates and target temperatures $T_i>T_f$. Because the overshoot temperature for the ThESE protocol is necessarily lower than $T_{\rm min}$ for the optimal protocol for a given $\Delta t$ --see Fig. \ref{fig1} (upper panel)-- the room temperature bound is reached by the ThESE protocol before the optimal one. More precisely, the ThESE protocol cannot accelerate cooling beyond $\Delta t/\tau= 0.6$, while remarkably and as perfectly measured, the optimal protocols can still have access to stronger acceleration rates with ratios between $\Delta t/\tau=0.6$ and $0.4$ that remain available experimentally. This important result reveals another, yet unexpected, thermodynamic advantage of optimization giving access to time-entropy regions that are simply forbidden to non-optimized protocols.

Contrasting with cooling, optimal heating protocols are not constrained by any fundamental limit (Fig.~\ref{fig2}, lower panel). With $\Delta \Sigma_{\rm sys}>0$ in Eq. (\ref{eq:Smin}), the production of entropy does not diverge and the system can be forced to thermalize arbitrarily fast. The optimal time-entropy bound for heating protocols is plotted in Fig. \ref{fig2} together with the experimental measurements obtained when implementing heating ThESE and optimal protocols.

\begin{figure}[htb!]
	\centering{
		\includegraphics[width=0.4\textwidth]{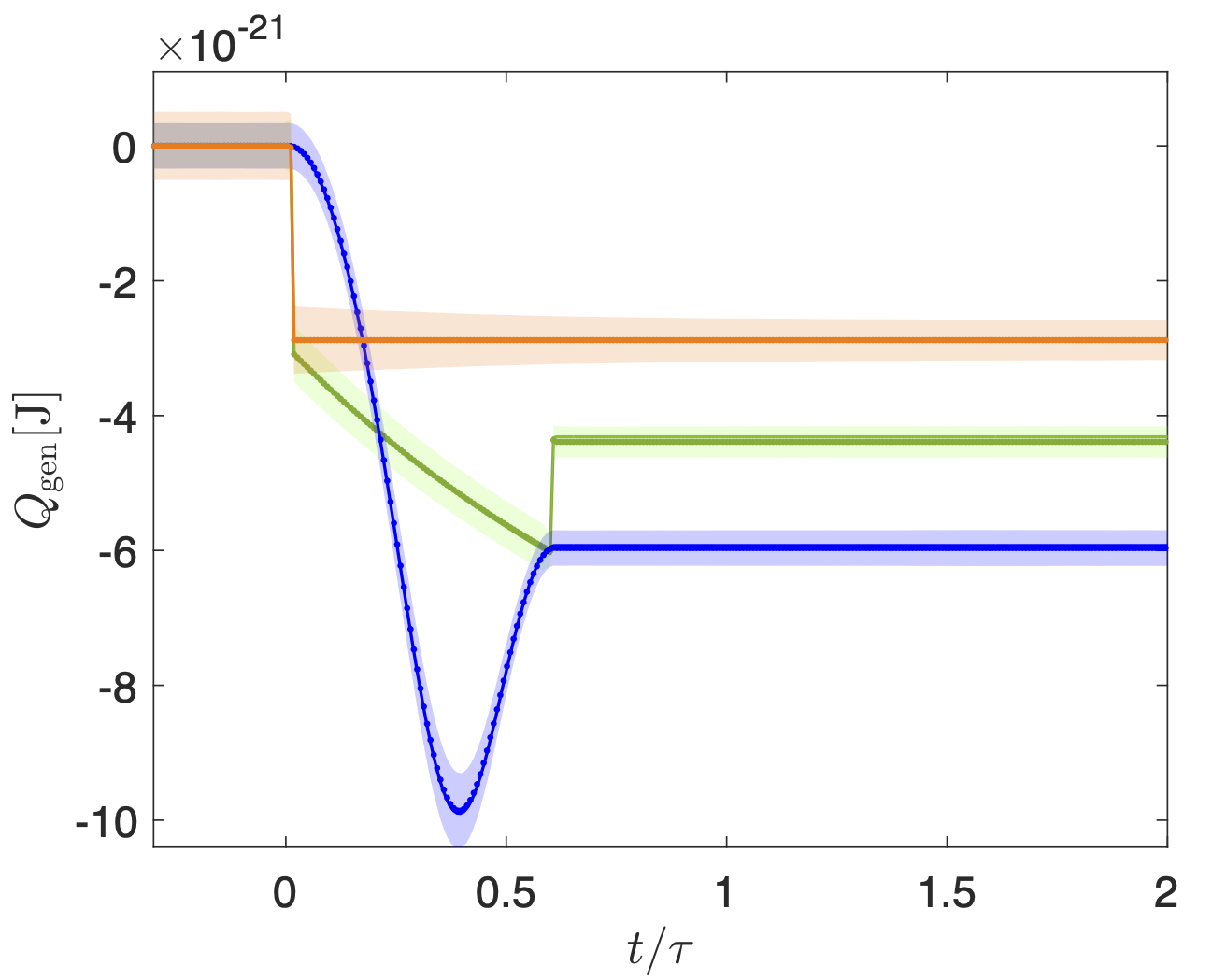}}
	\caption{Ensemble average, time-dependent cumulative generated heats $Q_{\rm gen}$ for a STEP (orange), a ThESE (blue) and an optimal protocol (green), all set at $t_i=0$ ms. Same color coding as in Fig. \ref{fig1} and same shaded areas associated with experimental errors at a $95\%$ confidence level, including calibration and temperature uncertainties evaluated by the expression of $Q_{\rm gen}(t)$.}
	\label{fig3}
\end{figure}

We finally measure the instantaneous, ensemble average, heat generated by the action of the radiation pressure $\dbar Q_{gen}= -T(t)d\Sigma_{\rm gen}$ and transferred to the microsphere while evolving from the initial state $s(t_i)$ to one non-equilibrium state $s(t)$ \footnote{The minus sign in this convention means that for a generation of entropy $d\Sigma_{\rm gen}>0$, the energy flows from medium to the system and $\dbar Q_{gen}<0$, consistent with the standard convention of stochastic thermodynamics \cite{sekimoto1998langevin}}. The time-dependent cumulative in-take of heat for isochores can be evaluated directly from Eq. (\ref{eq:Sgen}) as $Q_{\rm gen}(t)=-\int_{t_i}^{t}dt'\dot{T}(t')[k_{\rm B}T(t')-\kappa s(t')]/(2T(t'))$. This quantity of heat is plotted in Fig. \ref{fig3} for the three families of protocols studied here --STEP, ThESE, and optimal-- fixing a shortening ratio of $0.6$ for the ThESE and the optimal protocols. The time evolution and the final amount of $Q_{\rm gen} (t)$ strongly depend on the type of protocol. The energetic cost of the STEP protocol is relatively low, but it requires a long thermalization time. When comparing the other two protocols that have the same $\Delta t$, it is clear that the optimal solution mitigates the energetic cost of the transfer when compared to the ThESE protocol.

In conclusion, we have used a fluctuating, white-noise, radiation pressure to emulate temperature protocols applied to an optically trapped microsphere and to extend the concept of engineered swift equilibration to thermal protocols. A central result was to identify the entropic cost of such non-equilibrium protocols. The trade-off between the state-to-state transfer duration and the entropic cost led us to the design of optimal cooling and heating protocols. We identified minimal time-entropy bounds for all possible shortcut strategies in harmonic potentials and derived speed limits on the transfer rates. An energetic analysis showed in addition how optimization yields the best thermodynamic compromise between acceleration and cost. This optimization is important in the context of  thermodynamic cycles and Brownian heat engines, and bears a fundamental appeal considering that the entropic cost can be described as a thermodynamic length \cite{shiraishiSpeed2018}. From this perspective, our optimal cooling and heating shortcuts $T_{\rm opt}(t)$ correspond to geodesics within an information geometry viewpoint that draws fascinating connections yet to be further explored \cite{ito2018stochastic,ito2022geometric}.

\section*{Acknowledgments}

This work is part of the Interdisciplinary Thematic Institute QMat of the University of Strasbourg, CNRS, and Inserm. It was supported by the following pro- grams: IdEx Unistra (ANR-10-IDEX-0002), SFRI STRATUS project (ANR-20-SFRI-0012), and USIAS (ANR-10-IDEX- 0002-02), under the framework of the French Investments for the Future Program. 

\newpage

\onecolumngrid

\appendix

\section{Experimental setup}
\label{app.setup}

\begin{figure}[b!]
\includegraphics[width=100mm]{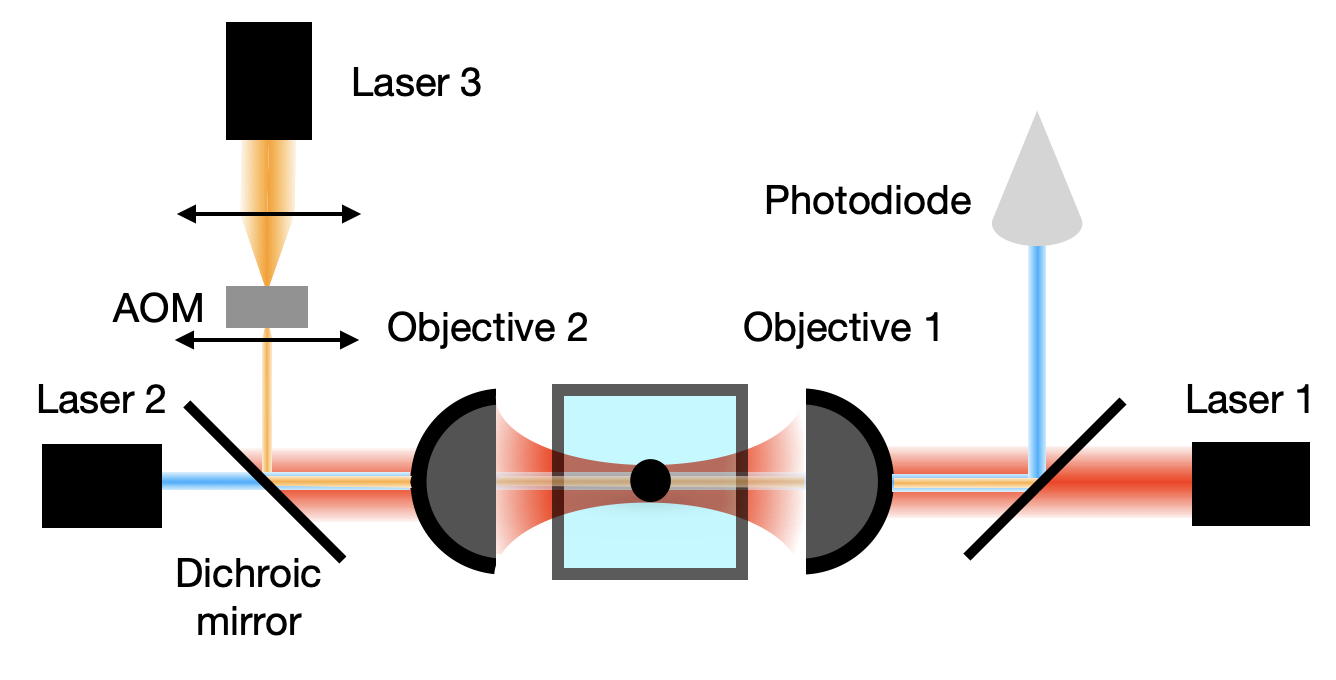}
\centering
\caption{\label{fig:setup}Experimental set-up. The trapping laser beam --laser $1$-- is focused inside the fluidic cell by objective $1$ ($60\times$, NA = $1.2$, water immersion). The instantaneous position of the trapped sphere is probed using laser $2$ whose scattered intensity is recorded using a P.I.N. photodiode. Laser $3$ exerts radiation pressure on the trapped sphere. Both lasers $2$ and $3$ are sent to the cell from the opposite side with respect to laser $1$ using objective $2$ ($60\times$, NA = $0.7$). It is important that neither laser $2$ nor laser $3$ induces any spurious gradient force inside the trap. For that, they are sent into objective $2$ in underfilling conditions. Laser $3$ passes through an acousto-optic modulator (AOM) --using the front lens of the  telescope stage to adjust laser waist and divergence to the AOM-- that enables driving in real-time its intensity (mean and fluctuations).}
\end{figure}

Our experimental setup consists in optically trapping a single polystyrene microsphere (Duke Scientific Corp. $3\ \mu \si m$ diameter) in ultra pure water by a Gaussian beam --laser $1$ on Fig. \ref{fig:setup} OBIS Coherent, CW $785\ \si{nm}$, $110\ \si mW $. The laser beam is expanded to overfill a high numerical aperture (NA) water immersion objective --objective $1$ Nikon Plan Apochomat $60\times$, NA = $1.2$-- that focuses the beam on a $18\ \si{\mu L}$ fluidic cell (glass slide and coverslip separeted by a spacer -Grace Bio-Labs) of thickness $120\ \si{\mu m}$. The microsphere solution is diluted to $0.5\times10^{-4}$ \% to ensure that only single spheres are trapped for all experiments. This is verified by looking to a bright field image produced by an additional laser -- not shown -- focused on the back focal plane of a low  NA objective --objective $2$ Nikon Plan-fluo extra large working distance $60\times $, NA=$0.7$.

The instantaneous position of the trapped sphere is recorded from the signal scattered off the trapped sphere of a diode laser beam --laser $2$ Thorlabs HL6323MG CW $639\ \si{nm}$, $30\ \si{mW}$-- focused and injected into the trap by Objective $2$. The scattered light is coolected by objective $1$ and directed toward a P.I.N. photodiode (Thorlabs, model Det100A2). The signal, recorded in volts, is sent to a low noise amplifier (Stanford Research, SR560) that removes through a $0.3\ \si{Hz}$ high-pass filter the DC component of the signal. A $100\ \si{kHz}$ low-pas filter is used in addition, to prevent aliasing. Both filters are set at $6 \ \si{dB/oct}$. The signal acquisition is finally done using an analog-to-digital card (National Instrument, PCI-6251) with an acquisition rate of $32768\ \si{Hz}$.

A third laser beam --laser $3$ Ti:Sapphire Spectra Physics 3900S-- adjusted to $800\ \si{nm}$ is used to exert fluctuating radiation pressure on the trapped sphere and emulate a secondary thermal bath. To do so, the laser is sent through an acousto-optic modulator (AOM - $3200$ S Gooch \& Housego) that modulates in real-time the intensity of the first-order diffracted beam. This beam is transmitted through Objective $2$ in underfilling conditions to couple to the trapped microsphere. The AOM output is measured by another P.I.N. photodiode -- not shown -- (Thorlabs, model Det100A/M), with the same configuration of the previous one. The control of the AOM is done using a digital-to-analog card (National Instruments PXRe-6738) with a generation rate of $20 \ \si{kHz}$. 

The divergence of laser $3$ is important to control to have an efficient radiation pressure coupling between this beam and the trapped microsphere. Since the focal plane of objective $2$ is first positioned to have a high signal-to-noise ratio for the detection of the instantaneous motion of the trapped sphere using laser $2$, a telescope is used to fine-tune laser $3$'s divergence so that both couplings are efficiently maintained through the common Objective $2$ configuration.

\section{Temperature calibration \label{app.calibration} }

\begin{figure}[htb!]
	\centering{
		\includegraphics[width=0.4\textwidth]{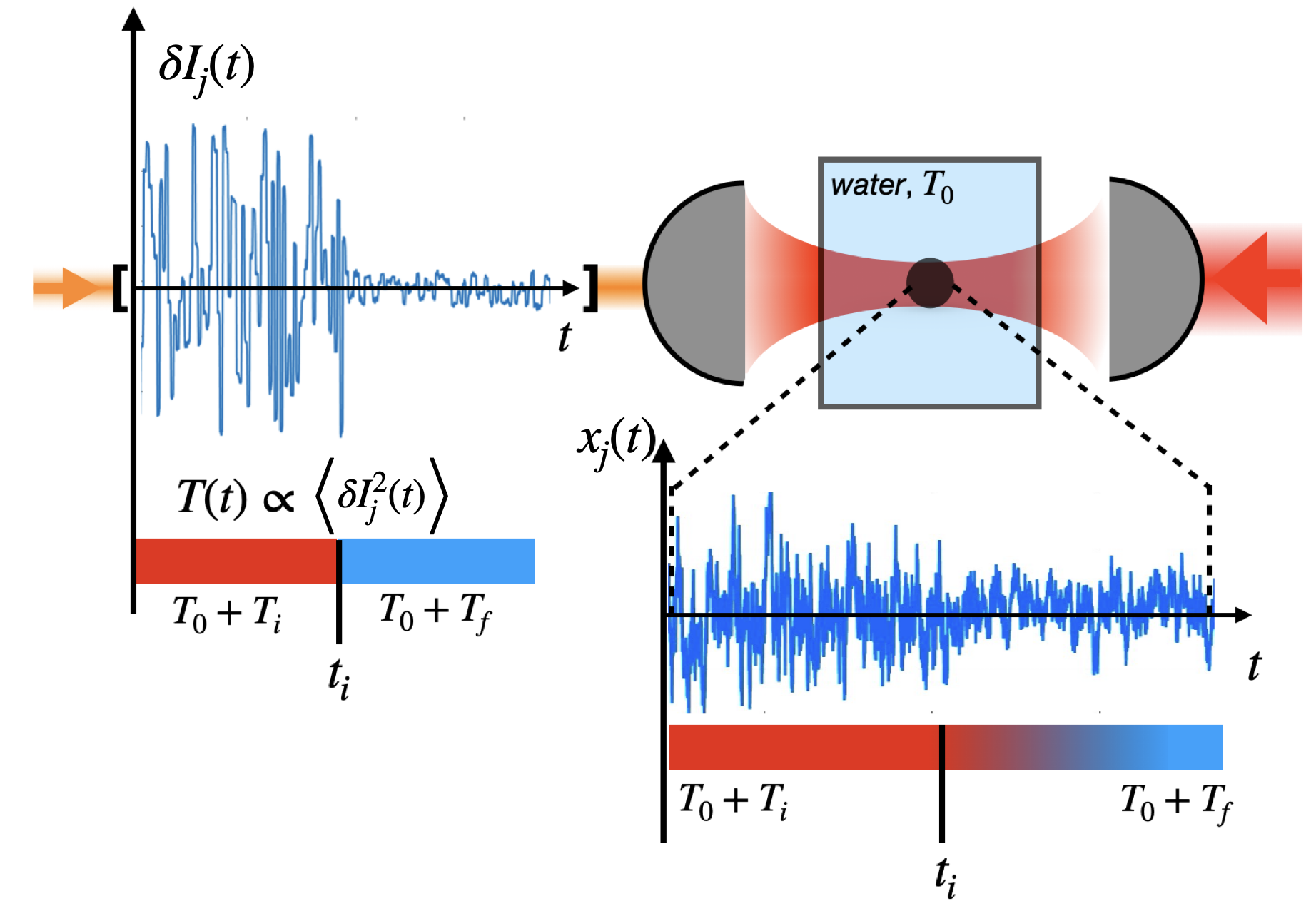}}
	\caption{A pushing laser, in orange on the left-hand side of the schematics (laser $3$ in Fig. \ref{fig:setup}), exerts unidirectional radiation pressure on the harmonically trapped microsphere and adds noise to the diffusing motion. As explained in the main text, one generates a step-like change in effective temperatures (here from hot --red-- to cold --blue) by modifying abruptly the variance of the laser white-noise intensity fluctuations at $t_i$. In contrast, the microsphere responds mechanically to this abrupt change by a transient evolution with a natural relaxation time $2 \tau$.}
	\label{fig:scheme}
\end{figure}

The intensity of laser $3$ is transformed, using the AOM, into a fluctuating signal with a white-noise spectrum. Exerting a white-noise fluctuating radiation pressure on the trapped microsphere, laser $3$ thus modifies the position variance $\langle x_j^2\rangle = s(t)$ of the microsphere, as measured over multiple trajectories $j$ ($1,7\times 10^4$ trajectories forming the experimental ensemble). More precisely, the laser instantaneous intensity is defined as $I_j = I_0 + \delta I_j$ for one trajectory. Ensemble averaging over the set ${j}$ of trajectories, we write $\langle I_j\rangle = I_0$ and $\langle(I_j-I_0)^2\rangle = \langle\delta I_j^2\rangle\propto \Delta T$. This radiation pressure emulates an effective thermal bath whose temperature can be changed instantaneously. The microsphere then thermalize within this effective, secondary bath, with variance values that depart from the equipartition set by room temperature, as schematized in Fig. \ref{fig:scheme}.

\begin{figure}[hbt!]
\includegraphics[width=120mm]{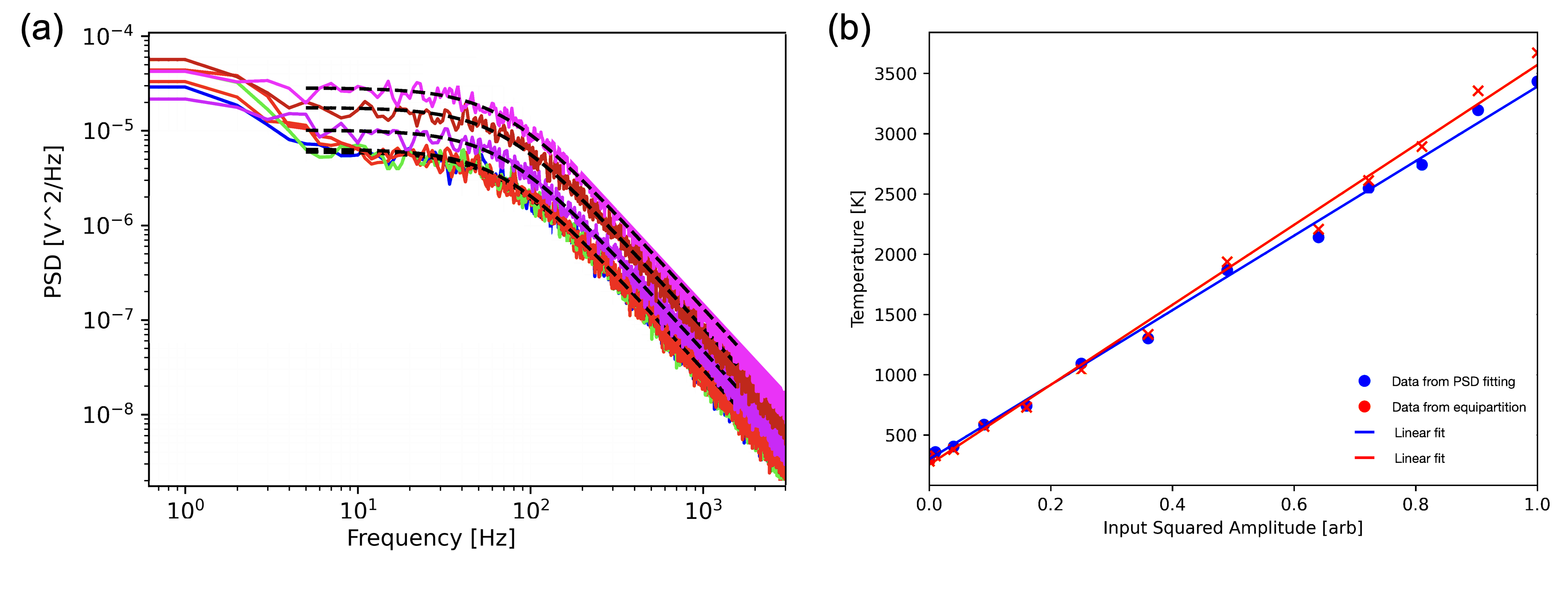}
\centering
\caption{(a) Position PSD for three non-fluctuating radiation pressure sets --lowest overlapping spectra-- and for three different white-noise intensity variances corresponding respectively to $T_{\rm R}+T= 500,\ 800, \ 1200\ \si{K}$. The dashed lines corresponds to Lorentzian fits of Eq. (\ref{eq.psd}) for finite bandwidths extending over the dashed fits. (b) Comparison between the two temperature calibration methods. The blue dots are the calibration data obtained by these Lorentzian fits and the red ones correspond to the temperatures obtained by assuming that $s$ obeys equipartition. The lines correspond to linear fits $T(\langle\delta I_j^{2}\rangle )= \alpha \langle\delta I_j^{2}\rangle + T_{\rm R}$.}
\label{fig:temp_calib}
\end{figure}

We make sure that the power spectrum of laser $3$ remains white over frequencies much larger than the roll-off frequency of the trap $f_c=2\pi / \tau$, and that the additional fluctuations imposed on the sphere do not affect the trap stiffness. This condition is met for sufficiently low intensities --$\sim 2$ mW-- so that, transmitted through objective $2$, laser $3$ does not impact the roll-off frequency of the Lorentzian position power spectrum (PSD) of the sphere set by the optical trap induced by Laser $1$ --see Fig. \ref{fig:temp_calib}. In the absence of fluctuations in the radiation pressure ($\langle\delta I^2_j\rangle = 0$), the natural PSD inside an optical trap of stiffness $\kappa$ is defined around the roll-off frequency $f_{c}=\kappa/(2\pi \gamma)$ as
\begin{equation}
\label{eq.psd}
S(f)= \frac{1}{2\pi^{2}\gamma} \frac{k_{B}T_{\rm R}}{f^{2}+f^{2}_{c}},
\end{equation}
where the viscosity of the water is taken at room temperature $T_{\rm R} = 293$ K with $\eta(T_R) = 9.532\times 10^{-4}\ \textrm{Pa}\times \textrm{s}$, the Stokes drag  $\gamma = 6\pi\eta r$ evaluated for a sphere radius $r = 1.5\ \mu\textrm{m}$. We calibrate the recorded voltage by fitting Eq. (\ref{eq.psd}) in the case of $\langle\delta I^2_j\rangle = 0$, following the standard procedure given in \cite{berg2004power,rosales2020optimal}.

The effective temperature associated with laser $3$ white-noise intensity fluctuations is measured from the evolution of the PSD --see Fig. \ref{fig:temp_calib}-- accounting for the limited operational bandwidth of the AOM. The temperature calibration exploits the linearity between the intensity variance of laser $3$ and the position variance of the trapped microsphere $s(t) = k_{\rm B}T_{\rm R}/\kappa + \alpha k_{\rm B}\langle\delta I^2_j(t)\rangle/\kappa$, for stationary conditions (or when the time dependence of $\langle\delta I^2_j(t)\rangle$ is slow enough to consider that $s(t)$ evolve as a succession of equilibrium states through which, at each time, $s(t)=k_{\rm B}T(t)/\kappa$ with $T(t)=T_{\rm R}+\alpha\langle\delta I^2_j(t)\rangle$).

The temperature calibration is performed using two different methods. The first one involves equipartition $\kappa s=k_{\rm B}(T_{\rm R}+T)$ between the steady-state, thermalized measured position variances $s$ and the effective temperatures $T$ set by laser $3$. This method assumes that high frequencies, those above the threshold imposed by the limited bandwidth of the AOM, do not contribute significantly to $s$. A second method that does not rely on this approximation is also implemented over a finite bandwidth analysis. For $\langle\delta I^2_j\rangle \ne 0$, we use Eq. (\ref{eq.psd}) to fit a Lorentzian on the position PSD, but now using the volt-to-meter conversion factor and the roll-off frequency value $f_{c}$ obtained in the case of a non fluctuating Laser $3$ intensity $I=I_{0}$. In that case, the only fitting parameter for the PSD is the effective temperature $T_{\rm R}+T$. The finite bandwidth extends over the dashed lines on Fig. \ref{fig:temp_calib} (a) in which the PSD for three different  $\langle\delta I^2_j\rangle$ that corresponds to $T_{\rm R}+T= 500,\ 800, \ 1200\ \si{K}$. The comparison between the two calibration methods is displayed in Fig. \ref{fig:temp_calib} (b) as a function of the input white-noise (squared amplitude) of laser $3$ fluctuations, $\langle\delta I_j^2\rangle$. The calibration factor $\alpha$ involved in the temperature changes $\Delta T =  \alpha\langle\delta I_j^2\rangle$ is then used in a \textsc{python} code \cite{goerlich2022harvesting} to build a generic $T(t)$ protocol applying a time dependent variance envelope $\langle\delta I_j^2 (t)\rangle$ in the noise signal sent by the AOM.

The uncertainties in stiffness and temperature are directly given by the calibration procedures. For the stiffness, uncertainties stem from the dispersion of the three measurements performed with their Lorentzian fits for $I=I_0$ and corresponds to $0.2$ fN/nm. For the temperature, uncertainty come from the errors in the fitting parameter involved in the second calibration method. They correspond to $ 40\ \si{K}$.

\section{Variance $s(t)$ equation of motion}
\label{APPENDIX_ExtNoise}

At the one trajectory $j$ level, the motion $x_j(t)$ of the trapped microsphere along the optical axis $\hat{x}$ of the optical trap follows the Ornstein-Uhlenbeck process
\begin{equation}
\label{eq:langevin}
\gamma\dot{x}_{j}(t)=-\kappa(x_j(t)-x_{0}(t))+F_{\rm ther}(\xi_{j}(t))+F_{\rm ext}(t),
\end{equation}
where, $F_{\rm ther}(\xi_{j}(t)) = \sqrt{2k_{\rm B}T_{\rm R}\gamma}\xi_{j}(t)$ is Gaussian distributed with zero average $\langle\xi_{j}(t)\rangle = 0$ and delta correlated $\langle\xi_{j}(t)\xi_{j^{\prime}}(t^\prime)\rangle= \delta_{jj^{\prime}}\delta(t-t^{\prime})$. 

The external radiation pressure produces a fluctuating force $F_{\rm ext}(t) = F_{0} + \delta F_j (t)$ with two contributions: the stochastic force contribution that acts on the Brownian particle and increases its variance, and the mean, constant contribution $F_{0}$ that produces a displacement in the equilibrium position. This spatial shift of the equilibrium position inside the trap can be taken into account by the following change of variable $x_j(t) \rightarrow x_j(t)+F_{0}/\kappa$, with the new variable obeying Eq. (\ref{eq:langevin}) with $x_{0}(t)\equiv 0$.

The other contribution $\delta F_j$ can be combined with $F_{\rm ther}$. When the fluctuation spectrum of the external radiation pressure is set to a white noise, $F_{\rm ext}(t)$ acts as a secondary, thermal bath according to $F_{\rm ther}(\xi_{j}(t))+\delta F_j (t)\rightarrow \sqrt{2k_{B}T(t)\gamma}\xi_{j}(t)$. This gives the possibility to perform kinetic temperature changes by simply adjusting the amplitude of such white-noise fluctuations. 

Multiplying Eq. (\ref{eq:langevin}) by $x_j(t)$ leads to the stochastic equation
\begin{equation}
\label{eq:langevin2}
\dot{x}_j^2(t)=-\frac{2}{\tau}x_j^2(t)+\sqrt{\frac{8k_{\rm B}T(t)}{\gamma}}\xi_j(t) x_j(t),
\end{equation}
where $x^2_j$ is experimentally accessible together with its ensemble average done over all stochastic trajectories $j$ at identical reference times $\langle x^2_j\rangle=s(t)$. Rather than given by an instantaneous derivative of the stochastic trajectory, $\dot{x}^2_j$ is given by the instantaneous difference $2x_j(t)(F_{\rm ther}/\gamma-x_j(t)/\tau)$. 

In order to transform Eq. (\ref{eq:langevin2}) in the deterministic variance equation evolution, we used the stochastic solution for $x_j(t)$ considering a generic $T(t)$ protocol
\begin{equation}
\label{eq:xjSolution}
x_{j}(t) = x_j(0^{-})\textrm{e}^{-t/\tau} + \int_0^t \sqrt{\frac{2k_{B}T(\zeta)}{\gamma}}\xi_{j}(\zeta) \exp{(\zeta-t)/\tau}\textrm{d}\zeta.
\end{equation}

From it, we evaluate the ensemble average correlation function between the stochastic trajectory and the effective thermal force $\langle x_j(t)F_{\rm ther}(\xi_j(t))\rangle = \sqrt{k_{\rm B}T(t)/2\gamma}$ and get Eq. ($1$), main text
\begin{equation}
\label{eq:var}
\frac{{\rm d}s}{{\rm d}t}= -\frac{2}{\tau} s(t) +2D(t),
\end{equation}
with the time dependent diffusion coefficient $D(t)= k_{\rm B}T(t)/\gamma$ and reminding that $\gamma = \tau\kappa$.

\section{Generation of entropy $\Sigma_{\rm gen}$ \label{app:Generation_of_entropy}}

Assuming that our Brownian sphere driven by a thermal protocol $T(t)$ evolves through a succession of equilibirum states, the probability density associated with the dynamical evolution can be written as: 
\begin{equation}
 \label{eq:prob1}
p(x_{j},T(t))= \frac{1}{\sqrt{2\pi k_{\rm B} T(t)/\kappa}}\exp{-\frac{\kappa x_j^2}{2 k_{\rm B} T(t)}}. 
 \end{equation}

Looking at the stochastic entropy $\sigma_{\rm sys} = -k_{\rm B}\ln p(x_{j},T(t))$ as a state function, the total variation of this quantity through a transition between two equilibrium states  $(\kappa,T_i)\rightarrow(\kappa, T(t))$ is protocol independent. This gives the possibility to extend Eq. (\ref{eq:prob1}) to the case of irreversible transformations for which the infinitesimal variation of $\sigma_{\rm sys}$ can be evaluated as:
\begin{equation}\label{eq:dS}
{\rm d}\sigma_{\rm sys}(x_j,T(t))=\frac{\kappa}{T(t)} x_j\textrm{d} x_j + \frac{k_{\rm B} }{2T(t)}\left( 1-\frac{ \kappa x_j^2}{k_{\rm B} T(t)}\right)\textrm{d}T.
\end{equation}

From the thermodynamic interpretation of Eq. (\ref{eq:langevin}) presented in \cite{sekimoto1998langevin}, the infinitesimal stochastic heat is defined as $ {\rm d}q=-\kappa x_j{\rm d}x_j$. We can thus identify the first term on the right-hand side of Eq. (\ref{eq:dS}) as (the opposite of) the infinitesimal variation of the medium entropy $\dbar \sigma_{\rm med} ={\rm d}q_{\rm qs}/T(t)$. The infinitesimal variation in the total entropy corresponds to the entropy generated through the irreversible transformation $\dbar \sigma_{\rm gen} = \dbar\sigma_{\rm med} + {\rm d}\sigma_{\rm sys}$. After an ensemble average among non equilibrium trajectories, it writes as $\langle \dbar \sigma_{\rm gen} \rangle=\dbar \Sigma_{\rm gen} $, so that the cumulative, ensemble average, generated entropy is:
 \begin{equation}
 \label{eq:sigmaGen}
 \Sigma_{\rm gen}(t)=  \frac{1}{2}\int_{t_{i}}^{t}\frac{\dot{T}(\zeta)}{T^{2}(\zeta)}\left( k_{\rm B} T(\zeta) -\kappa s(\zeta)\right) \ {\rm d} \zeta,
 \end{equation}
in which $\dot{T} = {\rm d}T/{\rm d}t$.

\section{\label{app:dimen} Temperature protocols}

In this section, the time dependent expressions for the three temperature protocols discussed in the main text (STEP, ThESE and optimal) are derived, together with the corresponding system's motional responses through the time-evolution of the variance $s(t)$. The expressions obtained here are plotted in Fig. $2$, main text.

\subsection{STEP protocols  \label{app:step}}
The STEP protocol corresponds to an abrupt temperature change. Using the step function for which $\Theta (t-t_{i}) = 0$ if $t< t_{i}$ and $\Theta (t-t_{i}) = 1$ if $t\ge t_{i}$, the protocol that connects an initial temperature $T(t_i^-)=T_{\rm R}+T_i$ to a final one $T(t_i^+)=T_{\rm R}+T_f$ in which $T_f-T_i = \Delta T$ writes as

\begin{equation}
\label{eq:Tstep}
T(t) = (T_{\rm R}+T_{i})+\Theta(t-t_{i})\Delta T ,
\end{equation}

Using this protocol into Eq. (\ref{eq:var}) --Eq. (1) in the main text-- and imposing an initial equilibrium condition $s(t_i)=k_{\rm B} (T_{\rm R}+T_i)/\kappa$ corresponding to equipartition, the evolution of $s(t)$ for $t>t_i$ is given by
\begin{equation}
s(t) =\frac{k_{B}}{\kappa}\left( T_{\rm R}+T_{f} -\Delta T \exp{-\frac{2}{\tau}(t-t_i)} \right).
\end{equation}

\subsection{ThESE protocols  \label{app:these}}

Adapting \cite{martinez2016engineered} (in which trap stiffness protocols $\kappa (t)$ are developed) to temperature protocols $T(t)$, we use the same third degree polynomial ansatz for the variance $s(t) = A t^{3} + B t^{2} + C t + D$, imposing initial $s(t_{i})=s_i=k_{\rm B}(T_{i}+T_{\rm R})/\kappa$ and final $s(t_{f})=s_f=k_{\rm B}(T_{f}+T_{\rm R})/\kappa$ equilibrium conditions together with $\dot{s}(t_{i})=\dot{s}(t_{f}) = 0$. We calculate the time-dependent variance $s(t)$ along the transition duration time $t_f-t_i=\Delta t$ as
\begin{equation}
\label{eq:sESE}
s(t) =\frac{k_{\rm B}}{\kappa}\left( -2 \Delta T \frac{ t^{3}}{\Delta t^{3}}  + 3\Delta T \frac{t^{2}}{\Delta t^{2}} + T_{\rm R}+T_{i} \right) .
\end{equation}

Substituting Eq. (\ref{eq:sESE}) and its derivative $\dot{s}(t)$ into Eq. (\ref{eq:var}) yields the explicit time dependent protocol
\begin{equation}
    T(t) = -2 \Delta T\frac{ t^{3}}{\Delta t^{3}} 
    + 3\Delta T \left(  -\frac{ \tau}{\Delta t} +1  \right)\frac{t^{2}}{\Delta t^{2}} + 3 \Delta T\frac{ \tau t }{\Delta t^{2}} + T_{\rm R}+T_{i}.
\end{equation}

\subsection{Optimal protocols \label{app:optimalprot}}

Optimal protocols are derived following the optimization method that we developed in \cite{rosales2020optimal} for trap stiffness protocols $\kappa (t)$. This method relies on treating the equilibrium state-to-state transfer duration and the energetic cost on an equal footing, with trade-off regulated by a Lagrange multiplier to built a functional that can then be minimized using standard Euler-Lagrange equations. In the case of trap stiffness protocols $\kappa (t)$, the energetic cost was identified as the dissipative work \cite{rosales2020optimal}. As discussed in the main text, for the case of thermal protocol that correspond to isochoric transition (i.e. work-free), the thermodynamic footprint of the protocol is of entropic nature. The trade-off is then regulated between the transfer duration $\Delta t =t_f-t_i$ and the production of entropy. 

The expression of the generated entropy  $\Sigma_{\rm gen}$ given by Eq. (\ref{eq:sigmaGen}) --Eq. ($2$) in the main text-- can be integrated by parts:
\begin{equation}
\label{eq:sigmaT(s)}
\Sigma_{\rm gen}[s_i,s_f;T(s)] = \frac{k_{\rm B}}{2}\ln \frac{T_{\rm R}+T_f}{T_{\rm R}+T_i} + \frac{k_{\rm B}}{2}\left( \frac{\kappa s_f}{k_{\rm B} (T_{\rm R}+T_f)} -\frac{\kappa s_i}{k_{\rm B} (T_{\rm R}+T_i)}\right) - \frac{\kappa}{2} \int_{s_i}^{s_f}\frac{\textrm{d}{s}}{T(s)}.
\end{equation}

The first term is identified as the protocol independent total variation of the system's entropy $\Delta \Sigma_{\rm sys}$. The second one is zero when considering that the initial and final states are fixed to be at thermal equilibrium, obeying equipartition, just like for ThESE protocols. In contrast, the last term depends on the profile of the protocol and thus carries the entropic contribution of the non-equilibrium process. 

The duration $\Delta t = t_f-t_i$ for the $T_{\rm R}+T_i \rightarrow T_{\rm R}+T_f$ transfer can be written as a functional of the variance $s(t)$ using Eq. (\ref{eq:var}):
\begin{equation}
\Delta t = \int_{t_i}^{t_f} \textrm{d}t =\frac{1}{2}\int_{s_i}^{s_f}\frac{\gamma {\rm d}s}{k_{\rm B}T(s)-s\kappa}.
\end{equation}

The trade-off between entropy production and state-to-state transfer duration is then given by:
\begin{equation}
\label{eq.euler-lagrange}
\int_{s_{i}}^{s_{f}} L[s,T(s)]{\rm d}s  = \int_{s_{i}}^{s_{f}} \left(  \frac{\gamma}{k_{\rm B}T(s) -s\kappa} -  \frac{\lambda\kappa}{k_{\rm B}T(s)} \right)\textrm{d} s,
\end{equation}
where $\lambda/k_{\rm B}$ is a Lagrange multiplier. The Euler-Lagrange equation $\textrm{d}/\textrm{d}s(\partial L/\partial T')-\partial L/\partial T= 0$, with $T'\equiv \textrm{d}T/\textrm{d}s$, will lead to the second order polynomial equation
\begin{equation}
\left(1-\frac{\lambda}{\tau}\right)T^2+2\frac{\lambda}{\tau}\frac{\kappa s}{k_{\rm B}}T -\frac{\lambda}{\tau}\left(\frac{\kappa s}{k_{\rm B}} \right)^2 = 0.
\end{equation}
with two solutions that form the temperature protocols related either to a heating protocol or to a cooling one, refereed with the sub-index ``h'' and ``c'' respectively. The protocols can be written in terms of the variance according to
\begin{equation}
\label{eq:T_s}
T_{\rm h/c}(s) =\frac{\kappa s}{k_{\rm B}\left(1\mp \sqrt{\tau / \lambda_{\rm h/c}}\right)},
\end{equation}
with associated heating/cooling Lagrange multipliers. The quasi-static limit of a reversible transition corresponds to $\lambda_{\rm h/c}\rightarrow \infty$.

To be implemented, explicit time dependent solutions $T(t)$ for the protocol are necessary. For the variance,  time-dependent solutions are obtained by substituting Eq. (\ref{eq:T_s}) into Eq. (\ref{eq:var}) to give:
\begin{equation}
\label{eq:neg_s_t}
s_{\rm h/c}(t) = s_{i}\exp[-2\frac{(t-t_i)}{\tau}\left(1- \frac{1}{1\mp\sqrt{{\tau}/{\lambda_{\rm h/c}} }
}\right)].
\end{equation}

Another way to express those optimal solutions is through the transfer time $\Delta t=t_f-t_i$, considering that at the final time of the transfer, the system is at thermal equilibrium with  $s_{\rm h/c}(t_f) = k_{\rm B}(T_{\rm R}+T_f)/\kappa$. The relation between $\lambda_{\rm h/c}$ and $\Delta t$ is thus given by
\begin{equation}
\label{eq:lagrange_multiplier}
1\mp\sqrt{{\tau}/{\lambda_{\rm h/c}}} = \left(1+\frac{\tau}{2\Delta t}\ln\frac{T_{\rm R}+T_f}{T_{\rm R}+T_i}\right)^{-1},
\end{equation}

The asset of this parametrization is to lead to a similar expression for a heating or a cooling protocol for $T(t)$ and $s(t)$. By substituting Eq. (\ref{eq:lagrange_multiplier}) into Eqs. (\ref{eq:T_s}) and (\ref{eq:neg_s_t}), we end up respectively with:
\begin{equation}
\label{eq:TOpt}
T_{\rm opt}(s,\Delta t,T_i,T_f) = \frac{\kappa s}{k_{\rm B}}\left( 1+\frac{\tau}{2\Delta t}\ln\frac{T_{\rm R}+T_f}{T_{\rm R}+T_i} \right),
\end{equation}
\begin{equation}
\label{eq:sOpt}
s_{\rm opt}(t) = s_{i}\left(\frac{T_{\rm R}+T_{f}}{T_{\rm R}+T_{i}} \right)^{(t-t_i)/\Delta t},
\end{equation}

\section{Minimal entropy production  \label{app:minimal}}

Eq. ($5$) from the main text is derived by substituting the optimal protocol, Eq. (\ref{eq:sOpt}) and Eq. (\ref{eq:TOpt}), into Eq. (\ref{eq:sigmaT(s)}). This leads to the entropy produced throughout the isochoric transition of duration $\Delta t$ associated with an optimal protocol, that is the minimal amount of produced entropy. 

As discussed in the main text, the optimal protocol goes through three stages: two discontinuities at the beginning and at the end of the protocol, and an exponential, non-equilibrium, evolution in-between for $t_i^+\leq t\geq t_f^-$. The first discontinuity with $T(t_i^-)=T_{\rm R}+T_i$ and $T(t_i^+)= (T_{\rm R}+T_i)(1+\frac{\tau}{2\Delta t}\ln \frac{T_{\rm R}+T_f}{T_{\rm R}+T_i})$, with $s(t_i^-)=s_i$ and $s(t_i^+)=s_i$, has an entropy production of
\begin{equation}
\label{eq:sigmaMin1}
\Sigma_{\rm min}^{(1)} = \frac{k_{\rm B}}{2}\ln \left[1+\frac{\tau}{2\Delta t}\ln
\frac{T_{\rm R}+T_f}{T_{\rm R}+T_i} \right] +\frac{k_{\rm B}}{2}\left(\frac{1}{1+\frac{\tau}{2\Delta t}\ln
\frac{T_{\rm R}+T_f}{T_{\rm R}+T_i}} -1 \right).
\end{equation}
 
Through the second intermediate stage $T(t_i^+)= (T_{\rm R}+T_i)(1+\frac{\tau}{2\Delta t}\ln \frac{T_{\rm R}+T_f}{T_{\rm R}+T_i})$, $T(t_f^-) =(T_{\rm R}+T_f)(1+\frac{\tau}{2\Delta t}\ln \frac{T_{\rm R}+T_f}{T_{\rm R}+T_i})$, $s(t_i^+)=s_i$ and $s(t_f^-)=s_f$, the entropy production is
\begin{equation}
\label{eq:sigmaMin2}
\Sigma_{\rm min}^{(2)} = \frac{k_{\rm B}}{2}\ln\frac{T_{\rm R}+T_f}{T_{\rm R}+T_i}-\frac{k_{\rm B}\ln\frac{T_{\rm R}+T_f}{T_{\rm R}+T_i}}{2\left(1+\frac{\tau}{2\Delta t}\ln
\frac{T_{\rm R}+T_f}{T_{\rm R}+T_i} \right)}.
\end{equation}

The last stage corresponds to the second discontinuity with $T(t_f^-) = (T_{\rm R}+T_f)(1+\frac{\tau}{2\Delta t}\ln
\frac{T_{\rm R}+T_f}{T_{\rm R}+T_i})$, $T(t_f^+) = T_{\rm R}+T_f$, while $s(t_f^-) = s_f$ and $s(t_f^+)=s_f$. It leads to the entropy production
\begin{equation}
\label{eq:sigmaMin3}
\Sigma_{\rm min}^{(3)} = - \frac{k_{\rm B}}{2}\ln\left[1+\frac{\tau}{2\Delta t}\ln
\frac{T_{\rm R}+T_f}{T_{\rm R}+T_i}\right] +\frac{k_{\rm B}}{2}\left(1-\frac{1}{1+\frac{\tau}{2\Delta t}\ln
\frac{T_{\rm R}+T_f}{T_{\rm R}+T_i}} \right),
\end{equation} 
with $\Sigma_{\rm min}^{(3)}=-\Sigma_{\rm min}^{(1)}$. The minimal amount of entropy production for an optimized thermal protocol through a transfer duration $\Delta t$ is $\Sigma_{\rm min} = \Sigma_{\rm min}^{(1)}+\Sigma_{\rm min}^{(2)}+\Sigma_{\rm min}^{(3)} =\Sigma_{\rm min}^{(2)} $. Using the definition of the total variation of the system's entropy $\Delta \Sigma_{\rm sys}=(k_{\rm B}/2)\ln (T_{\rm R}+T_f)/(T_{\rm R}+T_i)$, Eq. (\ref{eq:sigmaMin2}) leads to Eq. ($5$) in the main text.

\section{Energetics \label{app:energetics}}

As introduced above Sec. \ref{app:Generation_of_entropy}, the heat identified on Eq. (\ref{eq:langevin}) corresponds to a total differential. After averaging over the ensemble of trajectories, $ {\rm d} Q=-(1/2)\kappa \left<{\rm d}x^2_j\right>$, the cumulative heat corresponds to:
\begin{equation}
Q(t) = -\frac{\kappa}{2}\int_{s(t_i)}^{s(t)}{\rm d}s.
\end{equation}
Through an isochoric transformation (work-free), this cumulative heat is equal to the system's internal energy change, $\Delta U_{\rm sys} = -Q(t)$. The first law however does not account for the heat, given from the bath to the system, generated through the irreversible isochoric transformation. This energetic footprint can be evaluated based on $\Sigma_{\rm gen}$, using its differential form from Eq. (\ref{eq:sigmaGen})
\begin{equation}
\dbar \Sigma_{\rm gen} = \frac{k_{\rm B}T - \kappa s}{T^2}{\rm d}T.
\end{equation} 
This expression leads to define the in-take heat $Q_{\rm gen}(t)$ evaluated in the main text. In the reversible, quasi-static limit, the internal energies of the system and the medium coincide at all times, leading to
$Q_{\rm gen}(t)\rightarrow 0$ with therefore $k_{\rm B}\Delta T(t)/2 \rightarrow\Delta U_{\rm sys}(t)$ as expected.

\section{Data analysis and error bar \label{app:error}}

Once the temperature and trap calibrations are performed, as discussed above --Sec. \ref{app.calibration}-- the different temperature protocols are defined by setting the initial $T_{i}$ and final $T_{f}$ target temperatures and fixing the state-to-state transfer duration $\Delta t$. Each protocol is repeated to form an ensemble of $N_{cycles}\sim 17000$ trajectories recorded over $6$ minutes. These trajectories are combined, using for a time reference the intensity variance envelope sent by the AOM (see Sec. \ref{app.setup}). The uncertainty on the experimental variances are computed using a $\chi^{2}$ law with $N_{cycles}-1$ degrees of freedom with a confidence interval of $95\ \%$. The uncertainties on the entropy and heat measurements are obtained by standard methods for the propagation of errors. 

\bibliography{Biblio}

\end{document}